\documentclass[epj,referee,epsfig]{svjour}
\usepackage{latexsym}
\usepackage{graphics}
\begin{document}
\title{Interactions Between Charged Rods Near Salty Surfaces}
\author{Rebecca Menes\inst{1} \ Niels Gr{\o}nbech-Jensen\inst{2} \and Phil A. Pincus\inst{1}}
 
%
\offprints{}          
\institute{Materials Research Laboratory, University of California,
Santa Barbara, California, 93106, USA
\and Department of Applied Science, University of California,
Davis, California 95616, USA \\
NERSC, Lawrence Berkeley National Laboratory, Berkeley, California 94720, USA}
\date{Received: date / Revised version: date}
%
\abstract{
Using both theoretical modeling and computer simulations we study a model system for 
DNA interactions in the vicinity of charged membranes. We focus on the polarization
of the mobile charges in the membranes due to the nearby charged rods (DNA) and the resulting 
screening of their fields and inter-rod interactions. We find, both within a Debye-H\"uckel
model and in Brownian dynamics simulations, that the confinement of the mobile charges to 
the surface leads to a qualitative reduction in their ability to screen the charged rods
to the degree that the fields and resulting interactions are not finite-ranged as in systems
including a bulk salt concentration, but rather decay algebraically and the screening effect
is more like an effective increase in the multipole moment of the charged rod.
\PACS{
      {PACS-key}{describing text of that key}   \and
      {PACS-key}{describing text of that key}
     } 
} 
\maketitle
\section{Introduction}
\label{intro}

Recent experiments on various systems containing DNA strands and charged 
surfaces have raised an interest in understanding how these surfaces 
screen the electrostatic fields of the DNA and the resulting interactions
between strands. The interest emanates from sources varying from understanding 
prokaryotic DNA replication\cite{prokaryotic}, to non-viral gene
therapy\cite{gene,safinya,bruinsma,helmut}, and even DNA chip  technology\cite{chips}.

In this paper we treat this problem using a
two-dimensional salt solution model\cite{derjaguin,tots,velazquez,our,sam} to account for the  charged surface
(membrane), while the DNA strands are  modeled as negatively charged rigid rods. We treat charge
neutral systems  where
the over all charging of the surface and the rod is zero, thus focusing either on overall 
neutrally charged mixed lipid fluid membranes\cite{mixed}, or highly charged surfaces 
to which the
counter-ions are strongly bound and therefore treated within a 
two-dimensional geometry. We neglect the possible dependence on
 dielectric properties of the components in the experimental systems 
mentioned above in order to simplify the theoretical picture. The effects of
the thickness and the dielectric properties of the layers will be
published elsewhere\cite{next}. 

We focus on very simple geometries in order to understand how the DNA strand and
surface charges interact and screen. The geometry is that of one, 
infinite,
salty surface decorated with either one or two DNA strands. We calculate the
charge distribution around the DNA and the resulting interaction between the 
two
strands.  We calculate the interaction assuming the DNA strands are slightly 
raised above the surface (See 
Fig.\ 1).  

The theory we use to obtain our analytical results is within the Debye-H\"uckel
approximation\cite{landau}: We minimize the free energy of the system with respect to the
charge densities and use the result in the Poisson equation (thus obtaining the
Poisson-Boltzmann ({\bf PB}) equation) which we linearize with respect to the
electrostatic potential. Solving this equation leads to the optimized self
consistent charge distribution which we can insert back into the free energy 
in order
to obtain the resulting interactions. In the next section we introduce the model 
and the
formal results, and in the following section we apply it to find the 
interactions
between two strands.  We 
compare these results with new simulations of two-dimensional salt solutions 
and
 we conclude with a discussion on the limits of applicability and relevance of this model.

\section{Model}
\label{sec:1}
The free energy of a system of fixed and mobile charges includes electrostatic
terms and entropic terms:
\begin{equation}
F_{elec}={1\over 2}{e^2\over\epsilon}\int\left( 
{\sigma(\vec r)\sigma(\vec r')\over|\vec 
r-\vec
r'|}+ {\Sigma(\vec r)\Sigma(\vec r')\over | \vec r-\vec r'|}+2 {\sigma(\vec
r)\Sigma(\vec r')\over | \vec r-\vec r'|}\right)d\vec r d\vec r',
\end{equation}
\begin{equation}
S=-\int\left(\sigma_+(\vec r)(\log(\sigma_+(\vec r)a_0)
-1)+
\sigma_-(\vec r)(\log(\sigma_-(\vec r)a_0)-1)\right)d\vec r.
\end{equation}

Here $\sigma_+$ and $\sigma_-$ are the number densities of the positive and negative mobile 
charges where the total mobile charge density is given by $e\sigma=e\sigma_+-e\sigma_-$,
and $e\Sigma=e\Sigma_+-e\Sigma_-$ are the fixed charge densities.

Minimizing the Grand Potential:
\begin{equation}
 G=F_{elec}-k_BT S-\int\mu\left(
\sigma_++\sigma_-\right)d\vec r,
\label{gibbs}
\end{equation}
 with respect to the
mobile charge densities, $\sigma_+$ and $\sigma_-$, yields:
\begin{equation}
\sigma_+(\vec r)=a_0^{-1}e^{-e\phi(\vec r)+\mu\over k_BT}\;\;\;\;\;
\sigma_-(\vec r)=a_0^{-1}e^{e\phi(\vec r)+\mu\over k_BT}.
\label{boltzmann}
\end{equation}
Here the chemical potential of the positive and negative charges, $\mu$, is
taken to be equal since we treat the thermodynamic limit of an infinite 
system where the average number of positive and negative charges are equal 
far from the rods, and 
\begin{equation}
\phi(\vec r)={e\over\epsilon}
\int\left({\sigma_+(\vec r')-\sigma_-
(\vec
r')+\Sigma_+(\vec r')-\Sigma_-(\vec r')\over |\vec r-\vec r'|}\right)d\vec r',
\label{phidefine}
\end{equation}
is the resulting electrostatic potential. Inserting these results into the 
free energy we find the formal expression:
\begin{equation}
G={1\over 2}\int\left(\Sigma(\vec r)-\sigma(\vec r)\right)
\phi(\vec r)d\vec r. 
\label{energy}
\end{equation}
In this expression we have dropped constant terms that do not depend 
on the exact geometry of the system since we are interested in how
the  energy depends on the distances between the charges objects.
We will use this expression in the next section to
calculate the interactions in this system. 
Note
that the mobile charge density, $\sigma$, enters with an opposite sign to what
one would have  naively guessed to be the interaction. This is due to the fact that
this term enters as an entropic contribution and therefore indicates
how the entropy has been reduced (and thus the free energy increased) due to 
the arrangement of the mobile charges around the fixed charges.  

By inserting the distributions of Eq.\ \ref{boltzmann} in the Poisson equation 
we get the Poisson Boltzmann equation\cite{sambook}:
\begin{equation}
\epsilon\nabla^2\phi(\vec r)=-4 e\pi\left(\!\!\!\left(\!\!e^{-e\phi(\vec
r)+\mu\over  k_BT}-
e^{e\phi(\vec r)+\mu\over k_BT}\right)\!\!a_0^{-1}\!+\!\Sigma(\vec r\!)\!\right)\!\delta(z).
\end{equation}
The $\delta$ function was introduced because the charges are confined to the 
surface
at $z=0$. Linearizing this equation yields the Debye-H\"uckel ({\bf DH}) 
equation\cite{sambook}:

\begin{equation}
\epsilon\nabla^2\phi(\vec r)= \left({1\over\lambda}\phi(\vec r)-4\pi e
\Sigma(\vec
r)\right)\delta(z)\;\;\;;\;\;\;{1\over\lambda}={8 e^2\pi e^{\mu\over k_BT }\over k_BT a_0}.
\label{DH}
\end{equation}

Solving Eq.\ \ref{DH} for a given fixed charge distribution 
$\Sigma(\vec r)$, yields the mobile charge distribution and electrostatic
potentials and fields. In our case the fixed charge is that of a uniformly 
charged stiff rod (model DNA). We 
first solve Eq.\ \ref{DH} for a fixed point charge $Q$,
({\it i.e.,} $\Sigma(\vec r)=Q \delta(\rho)\delta(z)$) and then, since the 
problem is linear, we integrate to find the corresponding solution for an 
infinitely long rod. 

The potential that solves Eq.\ \ref{DH} has two contributions: a singular 
part, $\phi_{sing}={Q\over\epsilon\sqrt{\rho^2+z^2}}$, which solves the 
equation for a single point charge: 
$\epsilon\nabla^2\phi_{sing}(\vec r)=-4 e\pi Q\delta(\rho)\delta(z)$. The second
contribution,$\psi$,
 arises from the
mobile charges on the surface  and must satisfy the remaining equation:
\begin{equation}
\epsilon\nabla^2\psi(\vec r)=
 {1\over\lambda}\left(\phi_{sing}+\psi\right)\delta(z).
\label{DH2}
\end{equation}
Eq.\ \ref{DH2} is the Laplace equation: $\nabla^2\psi=0$, with the special 
boundary condition at $z=0$:
\begin{equation}
{\partial\epsilon\psi(0_+)\over\partial z}-{\partial\epsilon\psi(0_-)\over\partial z}={e\over\lambda}\left(\psi+\phi_
{sing}\right)|_{z=0}.
\label{BC}
\end{equation}

We solve for $\psi$ with a family of solutions of the Laplace equation:
\begin{equation}
\psi(\vec r)=\int \alpha_q e^{-q|z|}J_0(q \rho)dq. 
\label{psi}
\end{equation}
Using the identity $\phi_{sing}={Q\over\epsilon\sqrt{\rho^2+z^2}}={Q\over\epsilon}\int
e^{-q|z|}J_0(q\rho)dq$,  the boundary condition 
(Eq.\ \ref{BC}) is easily satisfied with 
$\alpha_q=-{Q/\epsilon\over 2q\epsilon\lambda+1}$, and the total electrostatic potential of
the system is  given by:
\begin{equation}
\phi_{point}=\phi_{sing}+\psi=Q\int e^{-q |z|}J_0(q\rho){2 q\lambda\over 2\epsilon 
q\lambda+1}dq.
\label{pointpot}
\end{equation}

This potential can now be integrated over a line to give the potential of a 
charged rod on a salty surface:
\begin{equation}
\phi_{rod}=\int\phi_{point}(\vec r)dy=2\tau\int e^{-q |z|}{\cos(q x)\over q}
{2 q\lambda\over 2\epsilon q\lambda+1}dq,
\label{rodpot}
\end{equation}
where $\tau$ is the charge per unit length on the bare rod (in the case of 
DNA: $\tau\simeq{-e/ 1.7\AA}$). Within this model the resulting charge distribution on the
surface
 is found to be:
\begin{equation}
\sigma(x)=-{1\over 2\pi\lambda}\phi_{rod}=-{\tau\over\pi}\int
{\cos(q x) dq\over 2\epsilon q\lambda+1}.
\label{chargedistribution}
\end{equation}

\section{Interactions}
\label{sec:2}

In this section we calculate the effective interactions between the components
in the system. The surface charges, which distribute themselves around the fixed
charges, screen to some extent their fields and thus the direct interactions. 
However, the screening is not as effective as that of a three-dimensional 
salt solution where the exponential screening leads to a finite
ranged interaction. In the case of a two-dimensional salt, although reduced, the fields are still
long ranged\cite{derjaguin}. This can be seen when we take the limit of large distances from the rod and
calculate the fields resulting from  Eq.\ \ref{rodpot}\cite{our}:
\begin{equation}
E_x(x\gg\lambda,z=0)\simeq{8\epsilon^2\tau\lambda^2\over x^3}\;\;\;
E_z(x=0,z\gg\lambda)\simeq{2\epsilon\tau\lambda\over z^2}.
\label{fields}
\end{equation}
These effective fields are dipolar in nature rather then the usual 
$1/r$ term for a charged line. However, they are not exponentially screened.
(Close to the rod (distances $<\lambda$) the electrostatic potential is 
not screened and therefore is logarithmic as is the case for a bare charged rod.)

\subsection{Interaction between two neighboring rods}
\label{subsec:2.1}
In order to calculate the interaction energy between two charged rods 
adsorbed on a salty surface we use Eq.\ \ref{energy} for the free energy 
of the system with the charge distribution of two rods  separated by a distance
$D$. We make use of the fact that the DH equation (Eq.\ \ref{DH}) is linear so 
that we can solve for each rod separately and then superimpose the 
potentials and charge distributions of the combined system of both rods. In 
order to compare with the simulation results which will be described
in the following chapter, we have to modify the problem slightly so that the
rods are  at least slightly raised above the surface and thus do not interfere with the
charge distribution around each other. (This is a requirement of the simulated 
system in order to allow for ions to move from one side of the rod to the other.)
 For a rod raised by a small (compared with
$\lambda$) distance
$d$ above the surface, the amplitudes of the modes in $\psi$ are now modified to 
be $\alpha_q(d)=-{\tau e^{-d q}\over\epsilon(2 q\epsilon\lambda+1)}$. For 
rods that are close to each other ($D<\lambda$) we know that the fields will
lead to a logarithmic inter-rod interaction. However, the fields farther away
are more complicated and the resulting interaction is not obvious. 

The interaction 
can be calculated numerically as a function of distance between the two rods, 
however, when the rods are separated by a distance $D\gg\lambda$ the 
potential (Eq.\ \ref{rodpot}) and charge distribution, $\sigma$ (Eq.\ 
\ref{chargedistribution}), can be 
analytically approximated yielding a simple form for the interaction energy as a 
function of the distance $D$. Once this approximation is made it can be shown 
that $\phi(x,z=0)\sim 1/x^2$ and $\sigma(x)\sim 1/x^2$ and the integrals in 
Eq.\ \ref{energy} are simplified to yield the interaction per unit length
 of the rods as a function of distance:
\begin{equation}
G(D\gg\lambda)\simeq{8 (\tau\lambda)^2\over\epsilon D^2}\left(1+{3\over 4}
{d\over\lambda}+{1\over 4}\left({d\over\lambda}\right)^2\right).
\label{2rodenergy}
\end{equation}
Here $D$ is the inter-rod distance and $d$ is the distance between the rods 
and the charged surface (Fig.\ 1).

The interaction is similar to that of two rods of dipoles at a
distance $D$ apart. However, we can not say that the charge distribution is
actually dipolar since the field in the perpendicular direction ($E_z$) 
behaves differently.

In general one usually expects the interaction between charged objects in 
solution to be dominated by the osmotic pressure of the solute, in this case 
the two-dimensional salt solution. However, in this two-dimensional case the
imperfectly screened electrostatic interactions dominate over the osmotic 
pressure which decays more quickly as a function of $D$ ( $\Pi_{osmotic}
\simeq k_BT(\sigma_+(D/2)+\sigma_-(D/2))\simeq O(\phi^2)\simeq O(D^{-4})$.
Differentiating equation\ref{2rodenergy} leads to the correct two-dimensional 
pressure between the rods which is dominated by electrostatic contributions:
\begin{equation}
\Pi=-{\delta G\over\delta D}={16 (\tau\lambda)^2\over\epsilon D^3}\left(1+{3\over 4}
{d\over\lambda}+{1\over 4}\left({d\over\lambda}\right)^2\right)
\label{pressure}
\end{equation}

\section{Numerical Simulations}
\label{sec:3}
In order to verify the predictions of the above theory, we have performed
numerical Brownian dynamics simulations of the effective interactions
between charged stiff (infinitely long) rods (along the $y$ direction) above
(in the $z$ direction) a surface, $(x,y, z=0)$,
in which charged monovalent particles can move (see Fig.\ 1). The
simulations are performed at a temperature, $T=300K$, and with uniform
dielectric constant, $\epsilon=80$, simulating the continuum properties
of bulk water. Simulating a non-zero salt concentration in the surface, we
have applied periodic boundary conditions in the plane $(x,y)$, with a
periodicity of $(L_x,L_y)=(400,40)$, where length is normalized to
$r_0=1${\AA}.
The normalized long range interactions between rods in this partially
periodic system are thus given by \cite{Mashl_JCP_110},
\begin{equation}
U_{rr}(x,z)  = 
-\frac{\tau_1\tau_2}{\epsilon}\ln\left(2\cosh\left(\!2\pi\frac{d}{L_x}
\right)
-2\cos\left(2\pi\frac{D}{L_x}\right)\right),
\end{equation}
and the normalized interaction, $U_{cr}(x,z)$,
between a point charge and a rod is given by
replacing $\tau_2$ with $q/L_y$ in the above expression, $q$ being the
fractional charge of the point charge. Energy is here normalized to
$E_0 = e^2/4\pi\epsilon_0$. The corresponding interaction energy between
point particles in partially periodic media can be found in Ref.\
\cite{Jensen_MolPhys}. We further employ a short range repulsive interaction
potential (in units of kcal/mol) between ions in the plane:
\begin{eqnarray}
U_{LJ}(r) & = & \left\{\begin{array}{lcc}
4\varepsilon\left(\frac{\rho}{r}\right)^6\left[\left(\frac{\rho}{r}\right)^6-1\right]+\varepsilon
& , & r<2^{1/6} \\ 0 & , & otherwise\end{array} \right. \; 
\end{eqnarray}
with $\rho=4$ and $\varepsilon=0.01$.

Figure 2 shows the simulation results for two rods of charge density,
$\tau=e/2${\AA}, at a distance $d=10${\AA} above the surface,
and with a distance
$D$ between them. The simulations have been performed by initiating the
positions of the in-plane ions randomly and allowing them to equilibrate for
$>10^5$ time steps ($dt=0.005$ in normalized time units) before averaging the
mean forces over $>10^6$ steps.
We show the attractive mean forces between the rods and the surface (Fig.\ 2a)
as well as the repulsive mean force between the rods (Fig.\ 2b) for several
different ionic strengths of the
surface. The prevailing trend is that the mean force between the rods
decays as $D^{-3}$ for large $D$ and that the mean force between the rods and
the surface approaches the asymptotic (single rod limit) as $D^{-2}$ for
large $D$. This is in agreement with the predictions given by
Eq.\ \ref{2rodenergy} and \ref{pressure}.
It should be noted that given the periodic boundary conditions necessary for
simulating a non-zero concentration of salt in the surface, we do not have
complete freedom to consider the limit  $D\rightarrow\infty$. The largest
possible $D$ is given by $L_x/2$ (where all forces between the rods,
in the $x$ direction, are
zero by symmetry) and one should therefore only consider simulation results
for $D$ somewhat less than $L_x/2$ in order to obtain meaningful results
to be compared to the theory, which does not consider a periodic array of
particles and rods. The results show some signs of insufficient averaging
for large $D$. Clearly, this is due to the very small mean forces that we try
to evaluate, combined with the rather small number of simulated particles
from which the averages are generated. The small systems are necessitated
by the long range interactions which require all charged objects to interact
with all other charged objects, thereby increasing the simulation time by the
square of the number of charged objects simulated. Since the averaging only
gets linearly better with the number of particles, we are limited to small
systems if we want to perform calculations of the electrostatics carefully.
Even so, the simulations do overall agree with the predictions. In fact, the
two main predictions, namely that the inter-rod force and the rod-surface
force asymptotically behave as $D^{-3}$ and $D^{-2}$, respectively, seem
to be very robust results even though they were derived in the thermodynamic 
limit of a large number of overall salt ions in the plane. The three different cases shown in
figure 2 represent ($\circ$) only counter-ions to the rods; 
($\times$) 
only salt at
concentration $1/400${\AA}$^{-2}$ (charge neutrality is satisfied by evenly 
smearing the rod counter-charge on the surface); and
($\triangle$)
$\circ$ and 
$\times$
combined. Even the case where only counter-ions are present shows reasonable
agreement between simulations and theory. We have also verified that good
comparisons between simulations and theory hold for other values of $d$,
the distance between rods and surface. The general trend is that as $d$
increases, so does the distance, $D$, at which the mean forces approach the
predicted slopes shown in figure 2.

\section{Conclusions}
\label{sec:concl}

We have presented both theoretical and simulation results for the interactions 
between charged DNA-like rods near a salty surface.  We have focused on the 
 polarization of the surface charge distribution and how it affects the fields 
and resulting interaction between two such rods. Our main conclusion is that 
when the mobile charges are confined, as is the case in our treatment, to a 
two-dimensional surface, they do not exponentially screen the fields, and hence
the effective interaction between the rods is not finite-ranged. Both theory and 
simulations show consistent power law interactions both for the force 
between the rods and for the force they apply to the charged surface. 

Despite the fact that the limits of applicability of the theory and simulation are not the 
same: the theory is valid at the limit of a large number of ions, and
assumes an infinite  surface with just two rods, while the simulation is restricted, for the
reasons  mentioned in section \ref{sec:3}, to a small number of particles and periodic 
boundary conditions, we still find a finite region of inter-rod distances 
where the two agree fairly well. This agreement indicates that although the theory 
is approximate it is robust non the less.

The systems we treat in this paper may seem relatively artificial because  
the charges are confined to the surface and there is no, 
or very little, residual bulk salt in the surrounding water solution. In addition,
we do not take into account the effects of the non-zero thickness of lipid membranes 
and the relatively low (compared to the surrounding water) dielectric constant of the lipid.
However, despite these limitations, we can apply these results to some experimentally 
investigated systems. Specifically, in recent x-ray experiments\cite{safinya} that studied
DNA-Cationic lipid complexes the structures that were found were formed by layers of membranes
intercalated  by ordered DNA domains. Most of these experiments were performed in very low bulk
salt  concentration, and moreover, because the counter-ions were trapped between the membranes 
they were effectively restricted to two-dimensional space. (The lower dimensionality of the space
is``measured" relative to the fixed charged objects that polarize the surrounding solute. In this case 
we are studying the screening of the DNA and therefore the space available to the ions for 
redistributing themselves is effectively two-dimensional compared with the DNA,
despite the fact that the ions are almost point like in comparison with the $20\AA$ diameter of the 
DNA.) This experimental
limit may also be valid in biological layered structures such as the Golgi apparatus.  Although the
effects of  dielectric discontinuities in these electrostatic systems can be nontrivial (we treat
these  elsewhere \cite{next}) they do not change the main results for distances $D\gg\lambda$,
and we do not expect the power law forces to change in this regime.

We are grateful to Cyrus Safinya, Ilya Koltover and Gerard Wang for 
discussing with us their experimental results. We thank Helmut Schiessel, Dov Levine and Nily Dan for useful
discussions.
PP and RM acknowledge the partial support of the MRL Program of the National
Science Foundation under Award No. DMR99-72246 in addition to Awards 8-442490-21587 and 8-442490-21825.
NGJ acknowledges support by the Director, Office of Advanced Scientific
Computing Research, Division of Mathematical, Information, and
Computational Sciences of the U.S.\ Department of Energy under contract
number DE-AC03-76SF00098. 

\begin{figure}


\caption{Schematic of the model system of membrane and DNA strands}
\label{fig:1}      
\end{figure}

\begin{figure}
\caption{Mean forces between rods and surface (a) and between the two rods (b)
as a function of the inter-rod distance, $D$.
System parameters are: $L_x=400$, $L_y=40$, $\tau=e/1.7${\AA}, $d=10$,
$T=300K$. Distances are in units of {\AA}. Forces in figure (a) are relative
to the force at $D=L_x/2$. Results are shown for three different two dimensional
salt concentrations (see figure).}
\label{fig:2}       
\end{figure}

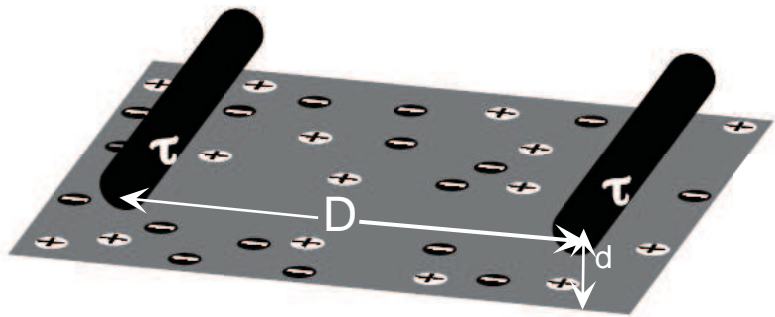
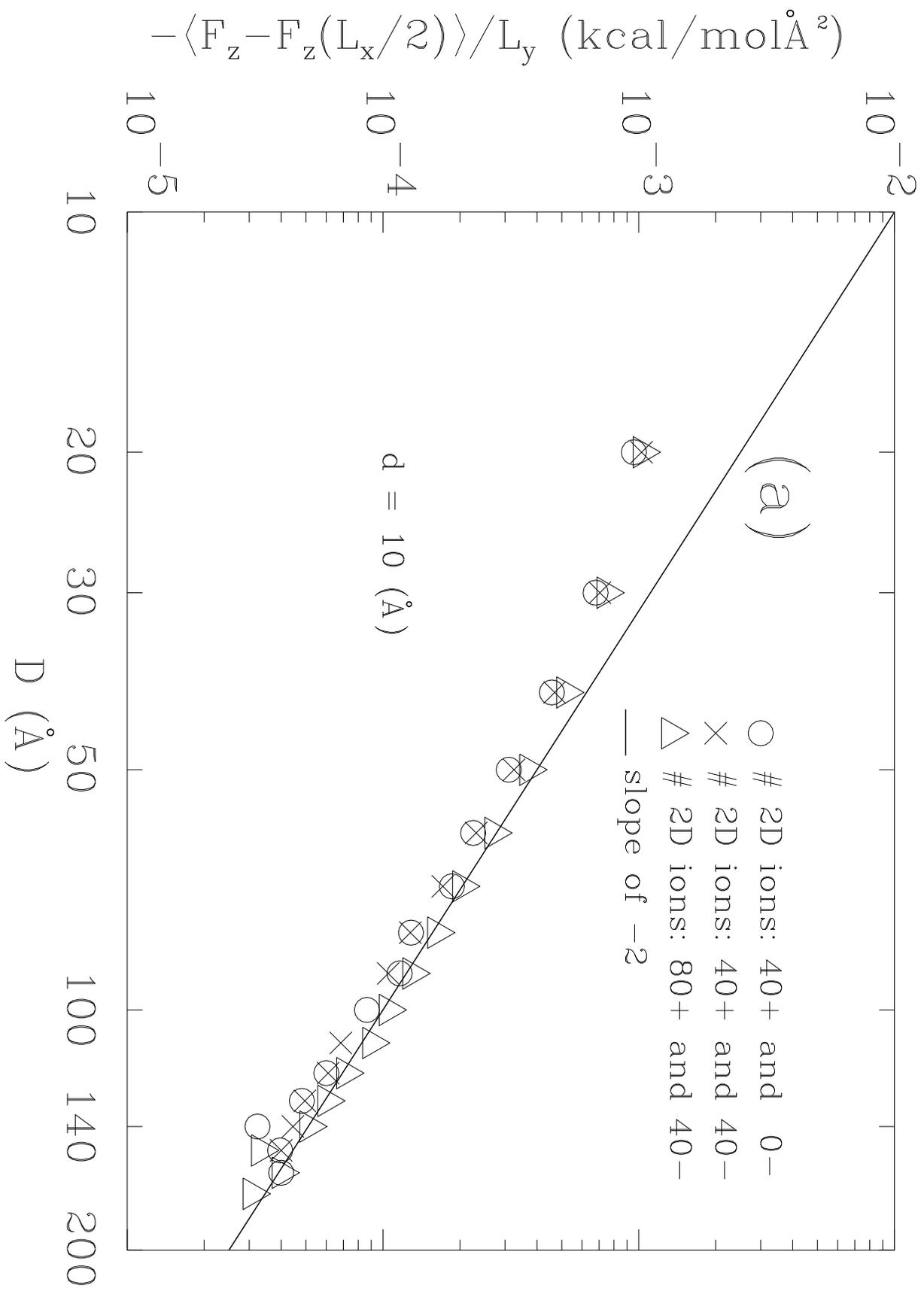
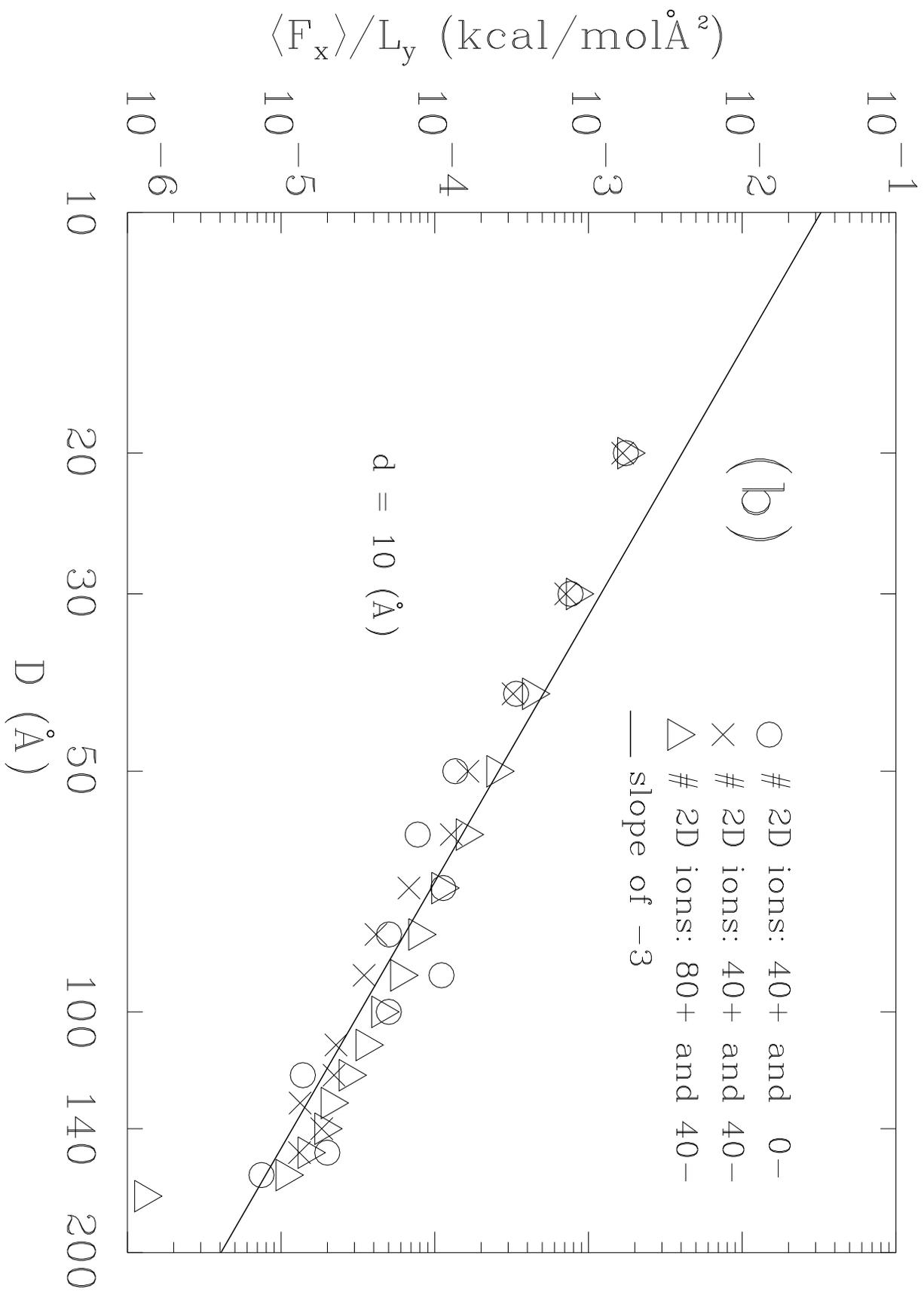
\end{document}